\pdfoutput=1
\documentclass[twocolumn,prb,showpacs,aps,superscriptaddress]{revtex4-1}

\usepackage[english]{babel}
\usepackage[ansinew]{inputenc}
\usepackage{times}
\usepackage{graphicx}
\usepackage{graphics}
\usepackage{amsmath}
\usepackage{amsfonts}
\usepackage{amssymb}
\usepackage{amsbsy}
\usepackage{dsfont}
\usepackage{epstopdf}
\usepackage{makeidx}
\usepackage{color} 
\usepackage{pgf}
\usepackage{bm}
\usepackage{tikz} 
\usepackage{graphicx}  
\usepackage{dcolumn}   
\usepackage{bm}        
\usepackage{amssymb}   
\usepackage{booktabs}
\newcommand{\be}{\begin{equation}}
\newcommand{\ee}{\end{equation}}
\newcommand{\bea}{\begin{eqnarray}}
\newcommand{\eea}{\end{eqnarray}}

\newcommand{\bu}{{\bf u}}
\newcommand{\bU}{{\bf U}}

\makeindex
\begin{document}

\title{Explicit Multi-element Extension of the Spectral Neighbor Analysis Potential for Chemically Complex Systems}
\email{athomps@sandia.gov}

\author{M. A. Cusentino}
\affiliation{Center for Computing Research, Sandia National Laboratories, Albuquerque, New Mexico 87185, USA}

\author{M. A. Wood}
\affiliation{Center for Computing Research, Sandia National Laboratories, Albuquerque, New Mexico 87185, USA}

\author{A. P. Thompson}
\affiliation{Center for Computing Research, Sandia National Laboratories, Albuquerque, New Mexico 87185, USA}

\date{\today}

\begin{abstract}
A natural extension of the descriptors used in the Spectral Neighbor Analysis Potential (SNAP) method is derived to treat atomic interactions in chemically complex systems.  
Atomic environment descriptors within SNAP are obtained from a basis function expansion of the weighted density of neighboring atoms.  
This new formulation instead partitions the neighbor density into partial densities for each chemical element, thus leading to explicit multi-element descriptors.  
For $N_{elem}$ chemical elements, the number of descriptors increases as $\mathcal{O}(N_{elem}^3)$, while the computational cost of the force calculation as implemented in LAMMPS is limited to $\mathcal{O}(N_{elem}^2)$ and the favorable linear scaling in the number of atoms is retained.  
We demonstrate these chemically aware descriptors by producing an interatomic potential for indium phosphide capable of capturing high-energy defects that result from radiation damage cascades. 
This new explicit multi-element SNAP method reproduces the relaxed defect formation energies with substantially greater accuracy than weighted-density SNAP, while retaining accurate representation of the bulk indium phosphide properties.  

\end{abstract}

\maketitle



%
\section{\label{sec:intro}Introduction}
Interatomic potentials (IAP) are an essential part of any classical molecular dynamics (MD) simulation, and are also the leading approximation determining the physical accuracy of this method.
The approximations that are inherent to all IAP (locality of forces, Born-Oppenheimer energy surface) are tolerated because this results in a computational cost that only scales linearly with the number of atoms, in contrast to cubic scaling of \emph{ab initio} MD.
Many IAP in usecapture local interactions using functional forms that approximate known physical and chemical phenomena, such as covalent bonding,\cite{Stillinger1985, Tersoff1988} electrostatic screening,\cite{Vashishta1990} electron density-mediated metallic bonding,\cite{Daw1983, Baskes1987} to name a few. 
In terms of developing new IAP, a decades long trend shows that much of the effort has focused on more accurate, but more computationally expensive potentials.\cite{Plimpton2012} 

A recent branch of this development incorporates advances in the field of data-science wherein machine learning of IAP can provide an alternative to the aforementioned physics inspired potentials.
Machine learned IAP (ML-IAP) take an alternate approach which is to forego a physically inspired model form in favor of a highly flexible functional based upon a generalized set of local atomic descriptors. 
These ML-IAP have an added requirement during parameterization relative to traditional IAP which is they need to be trained against a database of energies and forces that usually come from a higher fidelity simulation, e.g.  Density Functional Theory (DFT). 
The contributions from each of the chosen local descriptors to the energy and forces are independently weighted in order to match a database of higher fidelity results. 
A variety of different descriptors to describe the local environment exist in the literature,\cite{Bartok2013} such as symmetry functions\cite{Behler2007, Lindsey2017}, bispectrum components\cite{Bartok2010}, moment tensors\cite{Shapeev2016} and the Coulomb matrix.\cite{Rupp2012} 
At a minimum, a descriptor has to be invariant under translation, rotation and permutation, of neighboring atoms. 
As is the case in many machine learning applications beyond IAP, the development and selection of the descriptor space used is of critical importance to the overall model performance.  
Significant effort in recent years has focused on developing atomic descriptors that can be used in, though not limited to, material property predictions,\cite{ramprasad2017machine, kim2018polymer, nelson2013compressive, rajan2015materials, wang2019transferable, xie2019functional} alloy design,\cite{rosenbrock2019machine, nyshadham2019machine} and IAP used in MD.\cite{Behler2007, Bartok2010, Rupp2012, Bartok2013, Shapeev2016, Thompson2015}
Drautz has recently shown that many of these descriptors share a common mathematical foundation in the atomic cluster expansion for the Born-Oppenheimer potential energy function.\cite{Drautz2019}  
A recent study comparing families of ML-IAP trained on shared training databases using different descriptors and training techniques showed broadly similar improvements in accuracy with increasing number of descriptors.\cite{Zuo2020}
However, this comparative work by Zou \emph{et~al.} was limited to parameterizations of single element systems.
In this work we focus on the performance of the bispectrum components as atomic environment descriptors for multi-element systems. 

As is the case with all ML-IAP, SNAP reduction of regression errors (w.r.t. DFT training) can systematically be improved by including more descriptors, but Wood \emph{et~al.} showed there are diminishing returns in accuracy even though greater computational cost is incurred in calculating these extra descriptors.
These diminishing returns in accuracy hint that the underlying descriptor is incapable of capturing the full many-body nature of the energies and forces in the training set.  
In the original formulation of SNAP, the bispectrum descriptors only distinguish between neighbor atoms of different chemical elements based on their weighted contribution to the total atomic density.\cite{Wood2019,Li2018}  
This is similar in spirit to the construction of the density function within the embedded atom method for metal alloys.\cite{daw1993embedded,baskes1992modified}  
For systems that show strong differences in bonding characteristics depending on the chemical identity of the atoms, this weighted-density (WD) approach is likely insufficient.  

In this work we propose an explicit multi-element (EME) SNAP descriptor formulation.    
In EME-SNAP, the descriptors are decomposed into separate contributions that depend on the partial densities of each chemical element in order to better express differences in the chemical makeup of the local atomic environment.
In addition to the physical motivation to adapt these descriptors, the approach also adds more degrees of freedom to the model allowing for greater flexibility to fit the training data.
A derivation of the EME descriptors and a discussion of their implementation into the LAMMPS MD software is provided in Sections \ref{subsec:bispectrum} and \ref{subsec:snap} .  An example EME-SNAP potential and the associated training procedure is described in Section \ref{subsec:fitting} and a quantitative comparison of the accuracy of the WD-SNAP and EME-SNAP IAP is given in Section \ref{sec:results}.

\vline

\section{Computational Details}

\subsection{Explicit Multi-Element Bispectrum Descriptors}
\label{subsec:bispectrum}

In the original weighted density (WD) SNAP formulation, the total density of neighbor atoms around a central atom $i$ of element $\mu_i$ located at the origin is represented as a sum of $\delta$-functions in a three-dimensional space:
\begin{equation}
\rho ({\bf r}) = w^{self}_{\mu_i}\delta({\bf 0}) + \!\!\!\!\!\!  \sum_{r_{ik} < R_{cut}^{\mu_i\mu_k}} \!\!\!\!\!\!  {f_c(r_{ik}; R_{cut}^{\mu_i\mu_k}) w_{\mu_k} \delta({\bf r}_{ik})}
\label{eq:totaldensity}
\end{equation}

where ${\bf r}_{ik}$ is the position of the neighbor atom $k$ of element $\mu_k$ relative to the central atom $i$.  The $w_{\mu}$ coefficients are dimensionless weights that are chosen to distinguish atoms of different chemical elements $\mu$, while the central atom is assigned the weight $w^{self}_{\mu_i}$.  The sum is over all atoms $k$ within some cutoff distance $R_{cut}^{\mu_i\mu_k}$ that depends on the chemical identities of both the neighbor atom and the central atom.  The switching function $f_c(r; R_{cut}^{\mu_i\mu_k})$ ensures that the contribution of each neighbor atom goes smoothly to zero at $R_{cut}^{\mu_i\mu_k}$.  Following Bart{\'{o}}k \emph{et~al.},\cite{Bartok2010} the radial distance $r_{ik}$ is mapped to a third polar angle $\theta_0$ defined by,
\begin{equation}
\theta_0 = \theta_0^{max}\frac{r_{ik}}{R_{cut}^{\mu_i\mu_k}}
\end{equation}

The additional angle $\theta_0$ allows the set of points ${\bf r}_{ik}$ in the 3D ball of possible neighbor positions to be mapped on to the set of points $(\theta, \phi, \theta_0)$ on the unit 3-sphere.  
The neighbor density function can be expanded in the basis of 4D hyperspherical harmonic functions $\bU^j$
\begin{equation}
\rho({\bf r}) = \sum_{j=0,\frac{1}{2},\ldots}^{\infty}\bu_j\cdot\bU_j(\theta_0,\theta,\phi)
\end{equation}

where $\bu_j$ and $\bU_j$ are rank $(2j+1)$ complex square matrices.  The $\cdot$ symbol indicates the scalar product of the two matrices.  $\bu_j$ are Fourier expansion coefficients given by the inner product of the neighbor density with the basis functions $\bU_j$ of degree $j$.  Because the neighbor density is a weighted sum of $\delta$-functions, each expansion coefficient can be written as a sum over discrete values of the corresponding basis function, 
\begin{eqnarray}
\bu_j &=& w^{self}_{\mu_i}\bU_j({\bf 0}) + \\ 
\nonumber &&
\sum_{r_{ik} < R_{cut}^{\mu_i\mu_k}} {f_c(r_{ik}; R_{cut}^{\mu_i\mu_k}) w_{\mu_k} \bU_j(\theta_0,\theta,\phi)} 
\end{eqnarray}

The expansion coefficients $\bu_j$ are complex-valued and they are not directly useful as descriptors because they are not invariant under rotation of the polar coordinate frame.  However, the following scalar triple products of expansion coefficients are real-valued and invariant under rotation: \cite{Bartok2010}
\begin{equation}
B_{j_1j_2j}  = \\
 \frac{1}{2j+1} \bu_{j_1}\otimes_{j_1j_2j}\bu_{j_2}\cdot(\bu_j)^*
\end{equation}

The symbol $\otimes_{j_1j_2j}$ indicates a Clebsch-Gordan product of matrices of degrees $j_1$ and $j_2$ that produces a matrix of degree $j$, as defined in our original formulation of SNAP.\cite{Thompson2015}
The additional factor of $2j+1$ renders $B_{j_1j_2j}$ invariant under permutation of the indices, which simplifies the calculation of gradients (see Section~\ref{subsec:snap}).  These invariants are the components of the bispectrum.  They characterize the strength of density correlations at three points on the 3-sphere.  The lowest-order components describe the coarsest features of the density function, while higher-order components reflect finer detail.  The bispectrum components defined here have been shown to be closely related to the 4-body basis functions of the Atomic Cluster Expansion introduced by Drautz.\cite{Drautz2019}  

In the WD-SNAP method, the potential energy of each atom is written as a linear or quadratic function of these geometric descriptors, as described below. This has proven to be an accurate and efficient method for constructing interatomic potentials for both single element and multi-element systems.  However, because neighbor atoms of different elements are distinguished solely by the magnitude of the factor $f_c(r_{ik}; R_{cut}^{\mu_i\mu_k}) w_{\mu_k}$ that defines the effective weight of the contribution to $\bu_j$, the WD-SNAP formulation does not strongly distinguish the chemical identities of neighbor atoms.

In order to achieve a more explicit representation of different chemical elements, a natural step is to partition the total neighbor density into partial densities for each element
\begin{equation}
\label{eq:density}
\rho^{\mu} ({\bf r}) = w^{self}_{\mu_i\mu}\delta({\bf 0}) + \!\!\!\!\!\!  \sum_{r_{ik} < R_{cut}^{\mu_i\mu_k}} \!\!\!\!\!\!  {\delta_{\mu\mu_k} f_c(r_{ik}; R_{cut}^{\mu_i\mu}) w_{\mu} \delta({\bf r}_{ik})}
\end{equation}

where $\delta_{\mu\mu_k}$ indicates that only neighbor atoms of element $\mu$ contribute to the partial density $\rho^\mu$.  The central atom of element $\mu_i$ contributes a partial self-weight $w^{self}_{\mu_i\mu}$ to $\rho^\mu$.   By requiring that the partial densities sum to the total density used in the original WD-SNAP formulation
\begin{equation}
\rho({\bf r}) = \sum_{\mu=1}^{N_{elem}}{\rho^\mu ({\bf r})},
\end{equation}

it follows that $w^{self}_{\mu_i}$ in Eq.~(\ref{eq:totaldensity}) is equal to the sum of the partial self-weights $w^{self}_{\mu_i\mu}$.  In analogy with the WD-SNAP formulation, we set $w^{self}_{\mu_i\mu} = 1$ for all $\mu_i$ and $\mu$.

The partial expansion coefficients for each element follow naturally from this definition
\begin{eqnarray}
\bu^\mu_j &=& w^{self}_{\mu_i\mu}\bU_j({\bf 0}) + \\ \nonumber
&&\sum_{r_{ik} < R_{cut}^{\mu_i\mu_k}}  {\delta_{\mu\mu_k} f_c(r_{ik}; R_{cut}^{\mu_i\mu_k}) w_{\mu_k} \bU_j(\theta_0,\theta,\phi)} 
\end{eqnarray}

The explicit multi-element (EME) bispectrum can then be formed from products of the partial expansion coefficients
\begin{eqnarray}
B^{\kappa\lambda\mu}_{j_1j_2j}  &=& \frac{1}{2j+1} \bu^{\kappa}_{j_1}\otimes_{j_1j_2j}\bu^{\lambda}_{j_2}\cdot(\bu^\mu_j)^*
\end{eqnarray}

The EME bispectrum components are indexed on ordered triplets of elements.  Hence, for a two element system, each total bispectrum component is partitioned into eight EME bispectrum components.  In general, the bispectrum components are not invariant under permutation of the ordered triplet of elements.  However, when two or all three of the bispectrum indices are equal, then certain EME bispectrum components will be equal to each other.  These equivalences are expressed by the following identity
\begin{equation}
B^{\sigma(\kappa\lambda\mu)}_{j_1 j_2 j} = B^{\kappa\lambda\mu}_{j_1 j_2 j}, \quad {\rm \iff} \;\sigma(j_1,j_2,j) = (j_1,j_2,j)
\label{eqn:symm}
\end{equation}

where $\sigma$ is an element of the permutation group $S_3$.
The sum of the EME bispectrum components over all ordered triplets of elements is exactly equal to the total bispectrum component defined in the WD-SNAP formulation
\begin{equation}
B_{j_1 j_2 j} = \sum_{\kappa,\lambda,\mu=1}^{N_{elem}} B^{\kappa\lambda\mu}_{j_1 j_2 j}
\label{eqn:bsumrule}
\end{equation}

The EME bispectrum components defined in this way have a similar mathematical structure to the descriptors proposed by Drautz\cite{Drautz2019} in the multicomponent version of the Atomic Cluster Expansion (see Appendix A of Ref.~\citenum{Drautz2019}).

\begin{figure}[!t]
\includegraphics[]{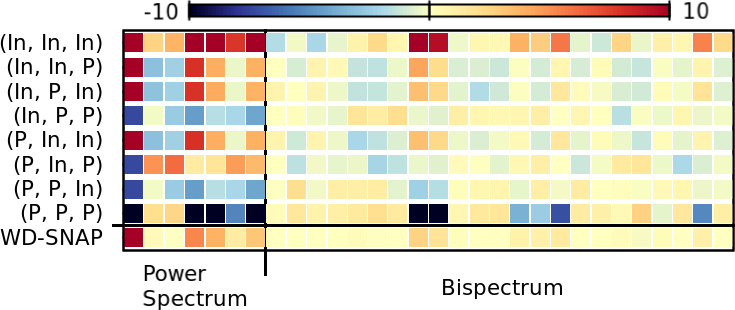}%
\caption{\label{fig:bispectrum} 
Difference in the bispectrum components between two very distinct chemical environments: the bulk phosphorous site and the phosphorous antisite defect.  The left panel shows the power spectrum components $(j,0,j)$, $0 \le 2j \le 2J_{max}=6$.  The right panel shows the other bispectrum components $(j_1,j_2,j)$, $0 < 2j_2 \le 2j_1 \le 2j \le 2J_{max}=6$.  The weighted-density SNAP descriptors are shown in the bottom row, while the upper rows show each of the explicit multi-element SNAP descriptors involving In and P.   
}
\end{figure}

In order to demonstrate the improved chemical sensitivity of the EME bispectrum components, we compare WD-SNAP and EME-SNAP descriptors for a phosphorous atom in two very different chemical environments.  In the zincblende ground state structure for bulk indium phosphide each phosphorous atom is covalently bonded to four indium neighbors.  In contrast to this, replacing an indium atom with a phosphorous atom creates an antisite defect in which the phosphorous atom has four phosphorous neighbors.  We calculated the EME bispectrum components for all 8 ordered triplets that can be formed from In and P, and for all half-integer triples $(j_1, j_2, j)$ in the range $0 \le 2j_2 \le 2j_1 \le 2j \le 2J_{max}=6$.  To simplify the comparison, we omitted the switching function $f_c$ and included only the four nearest neighbors of the phosphorous atoms by setting $R_{cut} = 4.1$~\AA.  In Fig.~\ref{fig:bispectrum} we show the difference between each of the descriptors in the two atomic environments.  It can be seen that the EME bispectrum components retain a large amount of extra information that is lost when they are summed up to form the total bispectrum components of WD-SNAP.  In particular, the (In, In, In) and (P, P, P) EME bispectrum components exhibit strong differences of opposite sign, which largely cancel out in the total bispectrum components.  This behavior is most pronounced in the case of the power spectrum components (left panel).

\subsection{EME-SNAP Potential Energy Function}
\label{subsec:snap}

Given the EME bispectrum components as descriptors of the neighborhood of each atom, it remains to express the potential energy of a configuration of $N$ atoms in terms of these descriptors.  
As in our previous work, we decompose the energy of the system containing $N$ atoms with positions ${\bf r}^N$ into a sum of local contributions that depend on the neighborhood of each atom and an additional reference energy $E_{ref}$ 
\begin{equation}
E({\bf r}^N) =  \sum_{i=1}^{N} E_i + E_{ref}({\bf r}^N)
\end{equation}

The reference energy is a convenient way to impose certain physical effects, such as long-range electrostatic interactions and strong short-range repulsion, for which well-established energy models exist.  Including a reference potential is advantageous because it can correctly represent known limiting cases of atomic interactions, leaving the many-body effects to SNAP.  The quality of the SNAP potential will somewhat depend on the choice of reference potential.  For example, since the training set does not include highly compressed configurations, the reference potential will need to provide a good physical description of Pauli repulsion which dominates the interaction at close separation.  In the case of the indium phosphide potential developed in this work, a previously described ZBL potential was used as the reference potential.\cite{Thompson2015, Ziegler1985}  The local SNAP contribution $E_i$ must capture all the additional effects that are not accounted for by the reference energy.  We assume that the local energy can be expressed as a linear function of all the distinct bispectrum components up to some maximum order $J_{max}$.  For a particular choice of $J_{max}$, we can list the $N_B$ total bispectrum components in some arbitrary order as ${B}_{1},\ldots,{B}_{N_B}$.  We can further decompose each  ${B}_l$ into EME bispectrum components $B_l^{\kappa\lambda\mu}$ and express the energy as a linear function of these
\begin{equation}
\label{eqn:snapatomenergy}
E_i({\bf B}_i)  = \sum_{\kappa,\lambda,\mu} \sum_{l=1}^{N_B} \beta_{l,\mu_i}^{\kappa\lambda\mu} 
(B_{l,i}^{\kappa\lambda\mu}-B_{l0,\mu_i}^{\kappa\lambda\mu})
\end{equation}

where $ \beta_{l,\mu_i}^{\kappa\lambda\mu}$ are the linear SNAP coefficients for atoms of element $\mu_i$.  
As a computational convenience, each EME bispectrum component is shifted by the value for an isolated atom, $B_{l0,\mu_i}^{\kappa\lambda\mu}$,
so that the local SNAP energy of an isolated atom is zero by construction.  The force on each atom $k$ of element $\mu_k$ is obtained by summing over all atoms of which it is a neighbor and all EME bispectrum components involving element $\mu_k$
\begin{eqnarray}
\label{eqn:snapforce2}
{\bf F}_k  &=& -  \sum_{i=1}^N \sum_{\kappa,\lambda}^{N_{elem}} \sum_{l=1}^{N_B} 
\beta_{l,\mu_i}^{\mu_k\kappa\lambda} \frac{\partial B_{l,i}^{\mu_k\kappa\lambda}}{\partial {\bf r}_k}+ \\ \nonumber
&& \beta_{l,\mu_i}^{\kappa\mu_k\lambda} \frac{\partial B_{l,i}^{\kappa\mu_k\lambda}}{\partial {\bf r}_k}+
\beta_{l,\mu_i}^{\kappa\lambda\mu_k} \frac{\partial B_{l,i}^{\kappa\lambda\mu_k}}{\partial {\bf r}_k}
\end{eqnarray}
The double sum over elements $\kappa$ and $\lambda$ in this expression shows that the computational cost of evaluating forces increases as $N_{elem}^2$.  The three terms correspond to the three different positions where element $\mu_k$ can appear in the chemical labelings of the EME bispectrum components.   

By formulating the SNAP potential energy as a linear function of the EME bispectrum components, the problem of generating the interatomic potential has been reduced to that of choosing the best values for the linear SNAP coefficients. We can achieve this by writing the SNAP contributions to the total energy, the force on an atom, and the stress tensor as explicit functions of the unknown SNAP coefficients ${\boldsymbol \beta_\mu}$
\begin{equation}
\label{eqn:snapenergy}
E({\bf r}^N) = \sum_{\mu=1}^{N_{elem}} {\boldsymbol \beta_\mu} \cdot \sum_{i \in \mu} {\bf B}_i
\end{equation}

where $ {\boldsymbol \beta_\mu}$ is the $(N_B \times N_{elem}^3)$-vector of SNAP coefficients for element $\mu$ and $ {\bf B}_i$ is the $(N_B \times N_{elem}^3)$-vector of EME bispectrum components for atom $i$.  The contribution of the SNAP energy to the force on atom $k$ can be written in terms of the derivatives of the EME bispectrum components w.r.t. ${\bf r}_k$, the position of atom $k$
\begin{equation}
\label{eqn:snapforce}
{\bf F}_k = - \sum_{\mu=1}^{N_{elem}} {\boldsymbol \beta_\mu} \cdot \sum_{i \in \mu} \frac{\partial {\bf B}_i}{\partial {\bf r}_k},
\end{equation}
where ${\bf F}_k$ is the force on atom $k$ due to the SNAP energy.  Finally, we can write the contribution of the SNAP energy to the stress tensor
\begin{equation}
\label{eqn:snapstress}
{\bf W} = -  \sum_{\mu=1}^{N_{elem}} {\boldsymbol \beta_\mu} \cdot \sum_{i \in \mu} \sum_{k=1}^{N}  {\bf r}_k \otimes \frac{\partial {\bf B}_i}{\partial {\bf r}_k}
\end{equation}
where ${\bf W}$ is the contribution of the SNAP energy to the stress tensor and $\otimes$ is the Cartesian outer product operator.  

All three of these expressions consist of the vectors $\boldsymbol\beta_\mu$ of SNAP coefficients for each element multiplying a vector of quantities that are calculated from the EME bispectrum components of atoms in a configuration.  This linear structure greatly simplifies the task of finding the best choice for $\boldsymbol\beta_\mu$.  We can define a system of linear equations whose solution corresponds to an optimal choice for $\boldsymbol\beta_\mu$, in that it minimizes the sum of square differences between the above expressions and the corresponding quantum results defined for a large number of different atomic configurations.  This is described in more detail in the following section.

\subsection{Fitting Procedure}
\label{subsec:fitting}
The process of fitting a ML-IAP has two key components, construction of a training set and optimization(i.e. learning) of free parameters in the model form. 
Both of these components will be detailed prior to the comparison of WD-SNAP and EME-SNAP descriptors.
In order to demonstrate the improvement over the WD-SNAP form of the EME-SNAP descriptors, a training set has been constructed for InP that exercises not only chemically unique environment but also contains high-energy defects which are challenging for all IAP to capture correctly. 
Training for accurate defect properties is critical when the intended use of the IAP is to study radiation damage effects where collision cascades of sufficiently high energy leave behind high formation energy point defects. 

Since the training of energies and forces for a ML-IAP is done against an electronic structure database, the training configurations are necessarily small ($N_{atoms}\lessapprox 10^{2}$) to make collecting a large number of them computationally tractable.
InP training configurations were generated using the Vienna Ab Initio Simulation Package (VASP)\cite{Kresse1993}$^{,}$\cite{Kresse1996a}$^{,}$\cite{Kresse1996b} using a LDA exchange correlation functional\cite{Perdew1981}, PAW pseudopetential\cite{Blochl1994}$^{,}$\cite{Kresse1999} leaving out the outermost valence p- and d- orbitals of P and In, respectively. 
K-point grids were generated using the Monkhorst-Pack\cite{Monkhorst1976} scheme with a spacing between 0.17 \AA$^{-1}$ and 0.72 \AA$^{-1}$ depending on system size, while a constant plane wave cutoff of $500 eV$ was used throughout.   

In total, 1206 configurations were generated with atom counts per configuration ranging from 8 to 216.
Curating training data for a SNAP potential involves grouping training data based on similarities between the configurations which limits the number of free parameters in the weighted least-squares regression.  
Each defect type(interstitials, vacancies, antisites and di-antisites) was assigned a unique group such that these configurations could be weighted individually during optimization.
In addition to defect configurations, we also included training groups to describe the properties of the bulk zincblende structure.  
This includes configurations for uniform expansion and compression (Equation of State), random cell shape modifications (Shear group), and uniaxially strained (Strain group) unit cells. 
Lastly, the ground state configuration for bulk zincblende structure is left as a separate group.  
The number of configurations($N_{config}$) in each training data group is summarized in Table~\ref{tab:groupWeights}.
Energies and forces are weighted separately, resulting in ten fitting degrees of freedom from the group weights. 
The weighting of different training groups is applied during the regression step within our fitting software FitSNAP\cite{fitsnapweb} by modifying the diagonal matrix $\bf w$ in the system of equations:
\begin{equation}
\label{eq:linreg}
{\min}(||{\bf w} (A \boldsymbol{x_{\beta}}- T)||^{2}-\gamma_{n}~||\boldsymbol{x_{\beta}}||^{n})
\end{equation}
Where $A$ is a matrix of bispectrum components computed from LAMMPS and $T$ is a vector of energies and forces taken from VASP. 
A regularization penalty, $\gamma_{n}$, can be used to aid against overfitting the solution $\boldsymbol{x_{\beta}}$, but to date has not yielded better fits than use without. 
 
In addition to weighting certain training data more/less heavily as part of the optimization, the radial cutoffs($R^{\mu_i\mu_k}_{cut}$ of Eq. \ref{eq:density}) and per-element density weights($w_{\beta}$ of Eq. \ref{eq:density}) are included as free parameters but are considered hyperparameters since they directly affect the calculation of the descriptors. 
All of these parameters have been described in previous work\cite{Thompson2015,Wood2019}, except for the atomic energy difference, which warrants further explanation.  
The DFT energy is computed as a sum of contributions from different terms in the full electron-ion quantum system, most of which are non-zero for an isolated atom.  
The SNAP potential, like most classical potentials, is constructed so that each isolated In or P atom has zero energy.  
To reconcile these two conventions, it is necessary to shift the DFT energy data by an amount that depends only on the element type of each atom, which we call the atomic energy. 
We constrained the sum of the In and P atomic energies by fixing the energy of the ground state structure at -3.48~eV/atom, ensuring that the experimental cohesive energy\cite{Harrison1989} of InP is reproduced.
The difference between the In and P atomic energies is an additional free parameter.  
  
\begin{figure}[!t]
\includegraphics{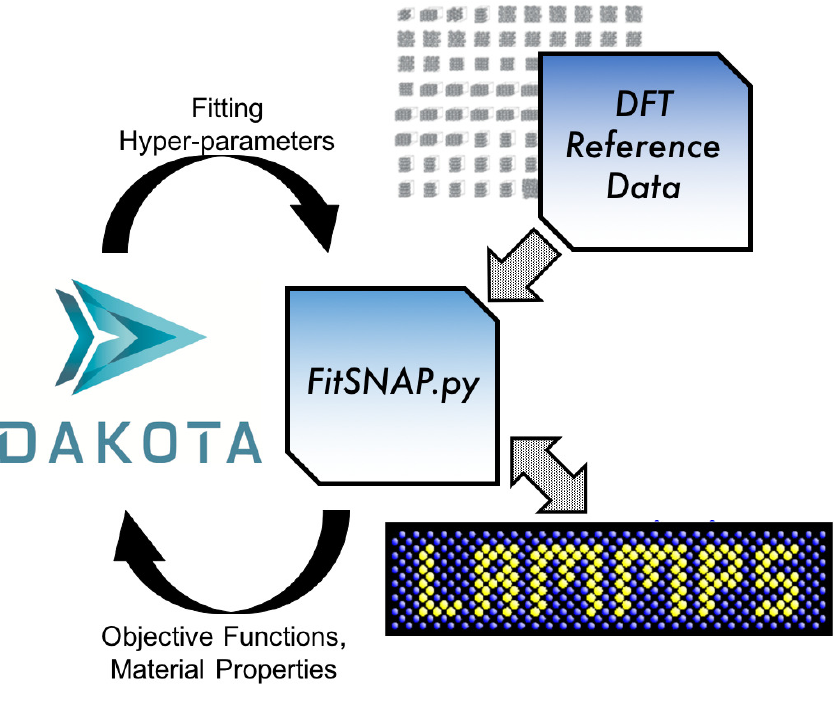}%
\caption{\label{fig:FitSNAP}
Schematic of the WD-SNAP and EME-SNAP fitting procedure.  The FitSNAP\cite{fitsnapweb} software package provides the overall workflow framework.  LAMMPS\cite{lammpsweb}efficiently evaluates the WD-SNAP and EME-SNAP descriptors, while DAKOTA\cite{dakota} performs the hyperparameter optimization.
}
\end{figure}

We automated the fitting process using the optimization software DAKOTA\cite{dakota}, the overall workflow is schematically captured in Figure \ref{fig:FitSNAP} and is the same when optimizing EME-SNAP or WD-SNAP ML-IAP.  
The hyperparameters and the group weights were optimized in two stages.  
First, the group weights were held fixed and the hyperparameters are optimized using a Single Objective Genetic Algorithm (SOGA). 
These hyperparameter search spaces were necessarily bounded, restricting the parameters to physical values and to save computational time, these ranges and the optimal values for each hyperparameter are given in Supplemental Table~I.  Once stable values for the hyperparameters were determined, a second SOGA optimization was performed on the training group weights which are allowed to vary over a large range($10^{1}-10^{7}$).  
 

For all DAKOTA driven optimizations, we have constructed a set of equally weighted objective functions that will be used to judge the quality of the fit.  
These objective functions are the lattice parameter and cohesive energy of zincblende InP as well as the difference between the DFT values and IAP predicted defect formation energies for the following stoichiometric pairs: In interstitial $+$ In vacancy, P interstitial $+$ P vacancy, In interstitial $+$ P interstitial, In vacancy $+$ P vacancy, In antisite $+$ P antisite.  
Referring back to the workflow shown in Fig.~\ref{fig:FitSNAP}, one pass through the optimization loop proceeds as follows.  
For each set of candidate hyperparameters or group weights proposed by DAKOTA, linear regression was used to solve for the SNAP coefficients using FitSNAP.\cite{SNAPSand}  
With each new candidate potential, LAMMPS is used \cite{Plimpton1995, lammpsweb} to relax a full set of InP defect configurations as well as the bulk InP zincblende structure.  
For each relaxation, the configuration was first annealed at 10 K for 0.1 ps before performing the minimization wherein the volume of the cell was also allowed to relax.  
The absolute error in these seven stoichiometric defect formation energies and the sum of the error in the cohesive energy and lattice parameter of zincblende formed the eight distinct SOGA objective functions.  
A generation consisted of 200 candidates after which hybridization and mutation steps that adjust group weights are carried out, the global fit was considered converged after 7600 generations.


\section{\label{sec:results}Results}
To demonstrate the improvement provided by the EME-SNAP descriptors optimized fits of either ML-IAP will be compared to available experimental data for InP properties near equilibrium as well as for high-energy point defects. 
For a complete comparison, we will also include predictions from a prior empirical IAP from Branicio \emph{et. al}.\cite{Branicio2009}
The Branicio potential for InP is a physics-motivated empirical model with electrostatic, van der Waals, and three-body interactions that take into account the covalent nature of the InP interaction.

To begin, Table \ref{table:alltrain} summarizes the errors for the reduced set of training groups outlined in the previous section.
For each training group and IAP type the mean absolute error (MAE) is reported for both the energy ($E_{MAE}$) and forces($F_{MAE}$).
In all cases except for the training group corresponding to sheared geometries, the EME-SNAP potential reproduces the DFT values of energy and atomic force most accurately. 
The average energy and force errors across the entire training set for EME-SNAP are 3.3~meV/atom and 67~meV/\AA, respectively, which are quite good for the training set size put forth here.
In contrast, the average energy and force errors for the SNAP potential are 5.7~meV/atom and 75~meV/\AA, respectively. 

\begin{table*}[]
\centering
\resizebox{\textwidth}{!}{%
\begin{tabular}{lcccccccccc}
\hline
\hline
&&&\multicolumn{2}{c}{Branicio}&\multicolumn{2}{c}{WD-SNAP}&\multicolumn{2}{c}{EME-SNAP}\\
Category&$N_{config}$&$N_{forces}$&$E_{MAE}$&$F_{MAE}$&$E_{MAE}$&$F_{MAE}$&$E_{MAE}$&$F_{MAE}$\\
\hline
Bulk&1&$2.4\cdot 10^1$&$4.8\cdot 10^{-4}$&-&$3.2\cdot 10^{-4}$&-&$5.8\cdot 10^{-4}$&-\\
Defects&428&$3.3\cdot 10^5$&$1.4\cdot 10^{-1}$&$3.0\cdot 10^{-1}$&$7.7\cdot 10^{-3}$&$1.5\cdot 10^{-1}$&$3.4\cdot 10^{-4}$&$8.5\cdot 10^{-2}$\\
EOS&268&$6.4\cdot 10^3$&2.5&$7.6\cdot 10^{-4}$&$1.1\cdot 10^{-2}$&-&$7.9\cdot 10^{-3}$&$8.2\cdot 10^{-4}$\\
Shear&346&$8.3\cdot 10^3$&$9.6\cdot 10^{-1}$&6.19&$3.5\cdot 10^{-3}$&$1.5\cdot 10^{-1}$&$4.6\cdot 10^{-3}$&$1.2\cdot 10^{0}$\\
Strain&163&$3.9\cdot 10^3$&$3.8\cdot 10^{-2}$&$3.6\cdot 10^{-5}$&$8.1\cdot 10^{-3}$&$2.4\cdot 10^{-5}$&$1.2\cdot 10^{-3}$&$2.4\cdot 10^{-5}$\\
\hline
\hline
\end{tabular}}
\caption{\label{tab:groupWeights}Mean absolute errors for energy and forces for the various groups of DFT training data for the Branicio,\cite{Branicio2009} WD-SNAP, and EME-SNAP potentials.}
\label{table:alltrain}
\end{table*}

\begin{figure}[!t]
\includegraphics[height=6cm,keepaspectratio]{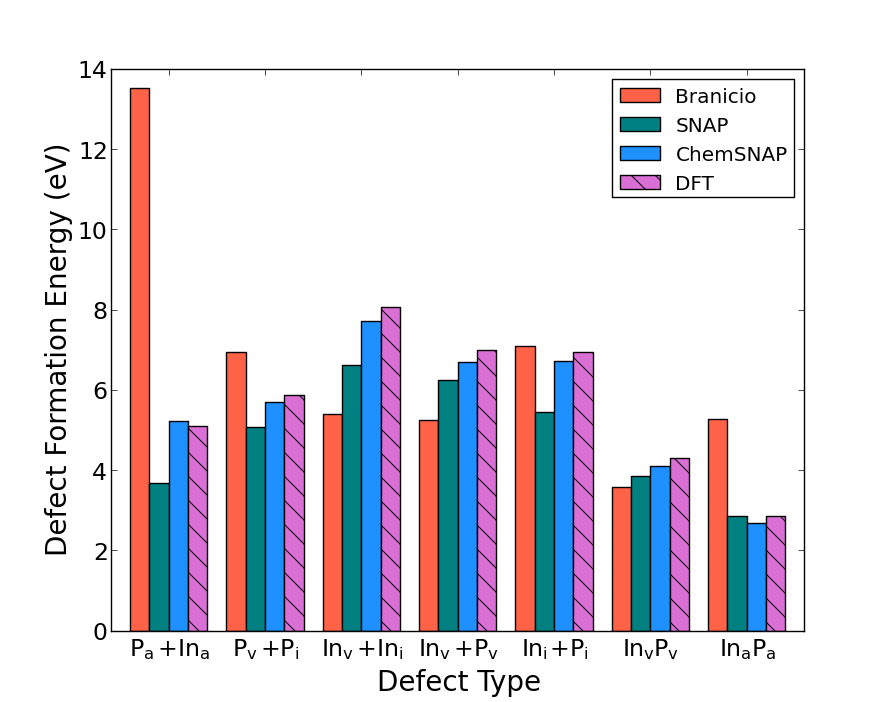}%
\caption{\label{fig:defectDFTChemSNAP} Relaxed defect formation energies for stoichiometric combinations of point defects in InP zincblende structure for classical potentials compared to DFT.  Orange, green, and blue represent the Branicio, SNAP, and EME-SNAP potentials, respectively.  The pink hatched columns represent the relaxed DFT formation energies.}
\end{figure}

Our main focus was to test whether EME-SNAP could reproduce the defect formation energies more effectively than WD-SNAP model.   
Fig.~\ref{fig:defectDFTChemSNAP} displays the defect formation energies for EME-SNAP (blue) and DFT (pink crosshatched).  
Subscripts in this Figure correspond to vacancy($v$), interstital($i$) and anti-site($a$) point defects.  As described previously, the defect formation energies are combined in stoichiometric combinations.  
Overall, EME-SNAP well reproduces the formation energies with the largest deviation from DFT being about 0.35 eV.  
With EME-SNAP, there are now eight descriptors for each $j$, $j_1$, and $j_2$ triplet combination(see Fig. \ref{fig:bispectrum}) and these descriptors can clearly distinguish between local environments comprised of varying compositions of multiple atomic species.  
This is especially useful for high-energy defects of lowered local symmetry compared to bulk InP. 
Importantly, the relative ordering of the seven formation energies is almost identical for DFT and EME-SNAP.  
The exception is that DFT predicts the combined In$_v$~$+$~P$_v$ formation energy to be about 0.06 eV higher than the combined In$_i$~$+$~P$_i$ defect, but EME-SNAP predicts the latter to be higher by 0.04 eV.  
This small difference will likely not drastically affect the defect population distributions obtained in radiation damage molecular dynamics simulations. 
The largest discrepancies between DFT and EME-SNAP are for the combined In$_v$~$+$~In$_i$ and the combined P$_v$~$+$~In$_v$, both are underpredicted by about 0.35 and 0.30 eV, respectively.  
In general, the interstitial configurations tend to be most difficult to accurately reproduce. 
However, this level of disagreement is acceptable, since the relative ordering of defects is accurately preserved.  
This is the most important factor in predicting the correct defect populations in collision cascade simulations and further evolution of the cascade.

Fig.~\ref{fig:defectDFTChemSNAP} also displays the relaxed defect formation energies for an empirical potential by Branicio {\it et~al.} \cite{Branicio2009} (red) and the WD-SNAP potential (blue). 
The Branicio IAP predicts defect formation energies poorly match DFT values with an average error of 2.46 eV, making this potential unsuitable for radiation damage studies.  
Regarding the ML-IAP fitted here, the WD-SNAP potential improves upon the Branicio potential but some of the defect formation energies still differ from DFT by as much as 1.4 eV.  
These discrepancies ultimately indicate the defects that would form during a collision cascade simulation using either the Branicio or WD-SNAP potential will most likely not be consistent with DFT.  
In contrast, the EME-SNAP defect formation energies are all within 0.35 eV of the DFT values and captures the overall trend of the DFT values.  
Many of the issues with the WD-SNAP potential stemmed from the configurations relaxing to a distinctly different structure during the minimization.  
In other words, the defect configuration predicted by DFT was only a metastable configuration for WD-SNAP.  
With EME-SNAP, the decrease in energy during MD relaxation is quite small and the structure predicted by DFT is generally preserved.  



\begin{table}[t]
\begin{ruledtabular}
\begin{tabular}{lccccc}
 &Branicio& WD-SNAP&EME-SNAP&Expt.&DFT \\
\hline
$a_0$ (\AA)&5.83&5.83&5.83&5.83&5.84\\
$C_{11}$ (GPa)&102.5&122.0&113.7&101.1&99.3\\
$C_{12}$ (GPa)&57.3&90.6&70.9&56.1&55.4\\
$C_{44}$ (GPa)&69.6&63.6&48.4&45.6&45.0\\
Bulk Mod.(GPa)&72.3&101.1&85.1&71.1&70.1\\
Shear Mod.(GPa)&22.6&15.7&21.4&22.5&21.9\\
Poisson Ratio&0.36&0.43&0.38&0.36&0.36
\end{tabular}
\end{ruledtabular}
\caption{\label{tab:elasticTable}Structural and mechanical properties of the InP zincblende structure, as predicted by the Branicio,\cite{Branicio2009} WD-SNAP and EME-SNAP potentials, as well as experimental\cite{Nichols1980} and DFT values.}
\end{table}

In addition to defect formation energies, we also examined the IAP predictions for other bulk InP properties.  
Table \ref{tab:elasticTable} lists the zincblende structure elastic constants for each of the the potentials.  
Given the simplicity of these equilibrium properties, it is unsurprising that the elastic constants are reasonably consistent with both experimental and DFT values for all potentials.
Still, there still are tradeoffs in accuracy among each.  
The Poisson ratio, bulk and shear moduli of EME-SNAP differ from DFT predictions by  5.6\%,21.4\% and 2.3\%, whereas WD-SNAP and the Branicio potential predict (19.4\%, 44.2\%, 28.3\%) and (0.0\%, 3.1\%, 3.2\%), respectively in this triplet of important properties. 
Overall, EME-SNAP does reasonably well in reproducing the elastic constants predicted by both experiments and DFT, but not to the level that the Branicio potential could.  
However, these values could potentially be improved upon by including stress tensor training data and adding the DFT elastic constants as objective functions during the fitting process.

\begin{figure}[!t]
\includegraphics[]{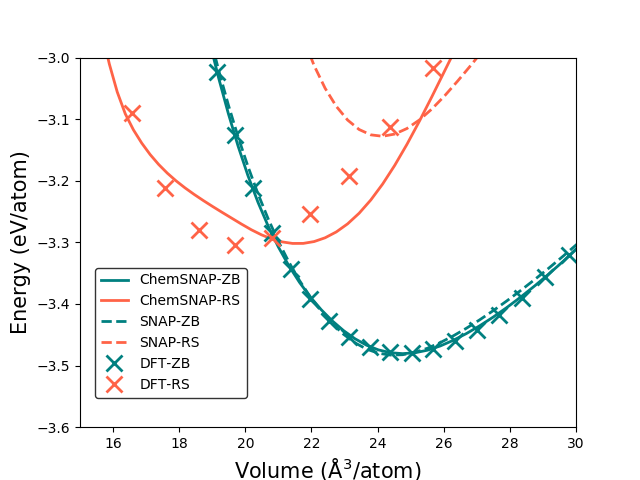}%
\caption{\label{fig:coldCurve} Energy versus volume for InP crystal structures calculated using the EME-SNAP potential (lines) and DFT (crosses).  Green and orange represent zinc-blende and rock salt structures, respectively.}
\end{figure}
 
Another important property is the relative stability of different low-energy crystal structures.  
Cold-curve equations of state for the zincblende(ZB) and rocksalt(RS) InP structures are compared with DFT and plotted in Fig.~\ref{fig:coldCurve}. 
EME-SNAP correctly predicts ZB as the most stable structure and reproduces the experimental cohesive energy of -3.48 eV/atom at a volume of 24.4\AA$^3$/atom while the RS cohesive energy of -3.30 eV/atom matches exactly the DFT value.  
However, EME-SNAP predicts a slightly higher volume of 22.2 \AA$^3$ compared to the DFT value of 19.69 \AA$^3$.  
While not plotted in Fig. \ref{fig:coldCurve}, the wurtzite(WZ) ground state structure was also calculated.  
The DFT prediction of the WZ cohesive energy was found to be -3.47 eV/atom, slightly higher than the ZB structure.  
While the EME-SNAP potential does predict the WZ to be higher in energy than ZB, the predicted value is too high, being about -2.27 eV/atom.  
The EME-SNAP representation of the RS and WZ properties is sufficient, given that these phases were not represented in the training data.  
Again, the accuracy of these phases could be improved in future iterations of this potential by including them in the training data and by using the RS, ZB and WZ cohesive energies as an objective functions during fitting.   

While the EME-SNAP InP potential significantly improves the defect formation energies and reproduces some basic InP properties, additional training data will be needed to further improve the transferability of the potential for other target applications.  
For instance, adding liquid phase training data would be useful as a liquid will sample a wide variety of local chemical environments.  
The overall fitting process could also be improved by incorporating other properties as objective functions.  
In this study, the defect formation energies were the primary focus, since accurately reproducing these properties is essential for realistic simulations of radiation damage, the focus of future work.  

\section{\label{sec:conclusions}Conclusions}
We have developed a natural extension to the SNAP interatomic potential form that improves the description of multicomponent systems.  
The new formulation, called EME-SNAP, uses chemically-labeled descriptors that explicitly separate out contributions from different chemical elements in the atomic environment.  
This new method was applied to InP where previous potentials were unable to adequately represent defect formation energies that are essential for conducting molecular dynamics simulations of radiation damage effects.  
The new EME-SNAP potential reproduced relaxed defect formation energies to within 0.35 eV compared to DFT, whereas the original weighted density SNAP formulation exhibited discrepancies of more than 1 eV.  
This improvement indicates that EME-SNAP is better able to distinguish between different chemical environments.  
Other properties including volume dependence of energy in the zincblende and rocksalt crystal structures are also well modeled by EME-SNAP.  
The new EME-SNAP method shows promise in generating a potential suitable for collision cascade simulations.  
The potential presented in this work was developed strictly for modeling defect formation energies and configurations close to equilibrium.  
It has yet to be examined whether this new formulation will perform well for radiation damage simulations.  
The training set will need to be expanded to incorporate additional configurations to create a more general use potential capable of simulating the full cascade.  
Future work will focus on using the EME-SNAP form to develop a potential with a broader training set and deploying it in radiation damage simulations.

\begin{acknowledgments}
The authors are grateful to Peter Schultz for constructive comments on an early draft of this paper. All authors acknowledge funding support is from the plasma surface interaction project of the Scientific Discovery through Advanced Computing (SciDAC) program, which is jointly sponsored by the Fusion Energy Sciences (FES) and the Advanced Scientific Computing Research (ASCR) programs within the U.S. Department of Energy Office of Science.
Equal support of this work is from the Exascale Computing Project (No. 17-SC-20-SC), a collaborative effort of the U.S. Department of Energy Office of Science and the National Nuclear Security Administration.
Sandia National Laboratories is a multi-mission laboratory managed and operated by National Technology and Engineering Solutions of Sandia, LLC, a wholly owned subsidiary of Honeywell International, Inc., for the U.S. Department of Energy's National Nuclear Security Administration under contract DE-NA0003525.  This paper describes objective technical results and analysis. Any subjective views or opinions that might be expressed in the paper do not necessarily represent the views of the U.S. Department of Energy or the United States Government.  
\end{acknowledgments}


\end{document}


\title{Supplemental Material }

{\let\newpage\relax\maketitle}


\begin{table*}[!t]
\caption{\label{tab:hyperParameters}Search ranges and final optimal values for EME-SNAP and WD-SNAP hyperparameters}
\begin{ruledtabular}
\begin{tabular}{lcccccccc}
&$w_{In}$&$w_{P}$&$R^{In}_{cut}$&$R^{P}_{cut}$&$E^{In}_{atom}+E^{P}_{atom}$&$E^{In}_{atom}-E^{P}_{atom}$\\
&(\textendash)&(\textendash)&(\AA)&(\AA)&(eV/atom)&(eV/atom)\\
\hline
EME-SNAP Search Range & \textendash &0.0 \textendash 1.5 &1.0 \textendash 5.0 & 1.0 \textendash 5.0 & \textendash &-100.0 \textendash 100.0\\
EME-SNAP Final Value &1.0&0.93&3.81&3.83&2.72&-6.04\\ 
\hline
\addlinespace[0.05cm]
&$w_{In}$&$w_{P}$&$R^{In}_{cut}$&$R^{P}_{cut}$&$E^{In}_{shift}$&$E_{shift}^{P}$\\
&(\textendash)&(\textendash)&(\AA)&(\AA)&(eV/atom)&(eV/atom)\\
\hline
WD-SNAP Search Range & 0.5 \textendash 1.5 & \textendash &1.0 \textendash 10.0 & 1.0 \textendash 10.0 & -100 \textendash 100 & -100 \textendash 100.0 \\
WD-SNAP Final Values&1.09&1.0&4.82&5.93&-1.19&1.09\\
\end{tabular}
\end{ruledtabular}
\end{table*}